\documentclass[
aps,
reprint,
prl,
superscriptaddress,
amsmath,
amssymb,
floatfix,
longbibliography,
]{revtex4-2}

\usepackage{bm}
\usepackage{graphicx}
\usepackage{latexsym}
\usepackage{array}
\usepackage{amsfonts}
\usepackage[free-standing-units]{siunitx}

\usepackage{mathtools}  
\usepackage{makecell}   
\usepackage{hhline}     
\usepackage{amsthm}
\usepackage{amscd}

\usepackage{xcolor}

\definecolor{orange}{rgb}{1,0.5,0}
\definecolor{goodGreen's}{rgb}{0.1,0.5,0}
\definecolor{goodred}{rgb}{0.7,0,0}

\usepackage[colorlinks,urlcolor=goodGreen's,citecolor=blue,linkcolor=goodred]{hyperref}

\newcommand{\orcid}[1]{\href{https://orcid.org/#1}{\includegraphics[width=8pt]{orcid.png}}}

\usepackage{verbatim}
\usepackage[normalem]{ulem}
\graphicspath{{images/}}

\usepackage{graphicx}
\usepackage{tikz}
\usetikzlibrary{arrows.meta, positioning, shapes.geometric, calc, decorations.markings}
\usepackage{amsmath}
\usepackage{amssymb}
\usepackage{mathtools}
\usepackage{bm}
\usepackage{xcolor}
\usepackage{booktabs}
\usepackage{multirow}
\usepackage{hyperref}

\definecolor{orbitA}{RGB}{12,68,124}
\definecolor{orbitB}{RGB}{121,31,31}
\definecolor{orbitC}{RGB}{39,80,10}
\definecolor{orbitD}{RGB}{99,56,6}
\definecolor{trsred}{RGB}{216,90,48}
\definecolor{invgreen}{RGB}{29,158,117}

\begin{document}


\title{Symmetry Classification of Non-Reciprocal Responses in Multiterminal Ring Devices}
\author{Chen-How Huang    }
\email{chenhow.huang@gmail.com}
\affiliation{Department of Physics and Nanoscience Center, University of Jyväskylä, P.O. Box 35 (YFL), FI-40014 University of Jyväskylä, Finland}
 \author{Tero T. Heikkil\"a }
 \affiliation{Department of Physics and Nanoscience Center, University of Jyväskylä, P.O. Box 35 (YFL), FI-40014 University of Jyväskylä, Finland}
\date{\today}
\begin{abstract}
We present a symmetry-based framework to classify the non-reciprocal responses of multiterminal ring quantum devices. The device is modeled as a ring of 
$n$ vertices, where a binary variable $e_k\in\{+1,-1\}$ on each bond encodes the preferred direction of signal flow between terminals. Non-reciprocity corresponds to a preferred current configuration on the ring, and the symmetry group of the device partitions all  $2^n$ configurations into equivalence classes(orbits) characterized by a topological winding number $W$. Using the minimal non-trivial case  
$n=3$, we establish two results independent of microscopic details. First, lifting the degeneracy within an orbit generates non-reciprocal responses. For 
$n=3$ this requires simultaneous breaking of both time-reversal 
$T$ and spatial inversion 
$I$. Breaking either alone is insufficient. Second, the residual geometry symmetry after $T$ and $I$ are broken determines which responses are observable. For an isosceles triangular geometry, only two types of response are allowed: uniform circulation (all bonds carrying current in the same direction) and semi-circulation with the reversed bond on the geometrically distinct base. Semi-circulation with the reversed bond on either equal leg is symmetry-forbidden. Both predictions are validated using a minimal toy model of three quantum dots coupled to superconducting baths, which demonstrates a reactive quantum circulator response. 
\end{abstract}
\maketitle
 \textit{\textcolor{blue}{Introduction---}}In recent years, a substantial paradigm shift has occurred, moving beyond the proof-of-concept stage toward the large-scale implementation of quantum computing technology. This progress is largely driven by the rapid evolution of circuit quantum electrodynamics (cQED), where the ability to precisely control and read out superconducting qubits has become the cornerstone of the field~\cite{SCQ_review_Kjaergaard2020,cQED_review_Blais2021}. As architectures scale toward higher qubit counts, increasing scientific emphasis has been placed on  microwave signal manipulation, specifically regarding the integration of non-reciprocal components and mesoscopic interfaces that facilitate high-fidelity, quantum-limited measurement \cite{diode_Ando2020,diode_Baumgartner2022,nr_PhysRevB.103.245302,3t_diode_Gupta2023,nr_Chiles2023}. Due to their practicality, the non-reciprocal quantum devices~\cite{LauClerk2018,Barzanjehetal2025,Hamannetal2018,cir_10.1063/5.0150427,cir_PhysRevX.5.041020,cir_Barzanjeh2017,cir_PhysRevX.7.011007} are the main-stream research in modern mesoscopic physics.  For two terminals, the only non-reciprocal response is forward versus reverse transport. As the number of terminals grows, multiple inequivalent non-reciprocal configurations become possible and it is no longer clear which responses are symmetry-allowed or how the device geometry selects among them. This motivates a systematic group-theoretic classification.
 
\begin{figure}[b]
\includegraphics[width=0.95\linewidth] {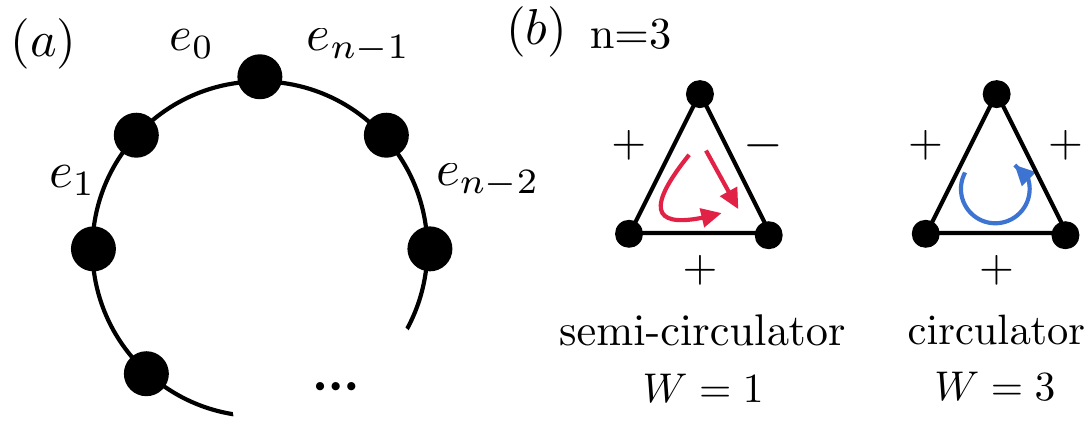}
\caption{ (a) $n$-vertex ring for the classification of non-reciprocal electronic response. Each bond is assigned a binary current $e_k\in\{1,-1\}$ describing the direction (counterclockwise, clockwise) of non-reciprocity. (b) $n=3$ ring of two different orbits classified by symmetries, the  semi-circulator and circulator configurations.}\label{fig:scheme} 
\end{figure}

 In this letter, we provide a generic theoretical framework based on group
theory to identify the possible non-reciprocal configurations of an 
$n$-terminal ring device and predict the allowed non-reciprocal responses
from symmetry alone, independent of microscopic details. We consider an $n$-terminal device in the form of a ring (Fig.~\ref{fig:scheme}). The terminals show up as nodes in the ring, connected via $n$ bonds, with the binary variable $e_k \in \{+1,-1\}$ indicating the preferred direction of signal flow in the bond. 
The symmetry group $G$ of the ring partitions all the $2^n$ current
configurations $\bm{e} = (e_0, e_1, \ldots, e_{n-1}) \in \{+1,-1\}^n \equiv \mathcal{C}$ into equivalence classes(orbits).
Non-reciprocity requires lifting the degeneracy between configurations in the same orbits which demands simultaneous breaking of time-reversal $T$ and spatial inversion $I$ at $n=3$.
Crucially, the symmetries that survive
after $T$ and $I$ are broken determine which non-reciprocal response
can actually be observed. We show two possibilities in Fig.~\ref{fig:scheme}, corresponding to circulatory and semi-circulatory non-reciprocity and indicate how this classification can be extended to generic $n$-terminal devices.

We apply this classification to the minimum non-trivial case $n=3$ and validate
the predictions using a minimal toy model of quantum dots coupled to superconducting terminals, demonstrating a reactive circulator or semi-circulator response depending on tuning conditions.

The macroscopic response of the ring can be characterized by the integer \emph{winding number}:
\begin{equation}
  W(\bm{e}) = \sum_{k=0}^{n-1} e_k \;\in\; \{-n,\,-n{+}2,\,\ldots,\,n{-}2,\,n\},
  \label{eq:winding}
\end{equation}
which counts the net circulation. The winding number 
$W$  cannot change without flipping at least one bond from $+1$ to $-1$. Configurations with different values of $W$ are therefore topologically distinct. For \emph{odd} $n$, $W$ is always odd and in particular $W = 0$ is topologically forbidden. For \emph{even} $n$, $W = 0$ is allowed and plays a special role. For instance, $W=0$ corresponds to the diode response in $n=2$ ring (see SM~\ref{app:n2}). Finally, although we focus on the ring geometry, the entire formalism can be applied to more complicated graph construction.

\textit{\textcolor{blue}{Symmetry generators and group structure---}}The symmetry group $G$ is generated by two main operations ($T$ and $I$) and the other geometrical operations of the system.  The time reversal and inversion operations are defined as, \textit{time-reversal $T$ (order 2).}
Reversal of the arrow of time flips all current directions:
\begin{equation}
  T:\;(e_0,e_1,\ldots,e_{n-1}) \;\mapsto\; (-e_0,-e_1,\ldots,-e_{n-1}).
  \label{eq:T}
\end{equation}
$T$ satisfies $W(T\bm{e}) = -W(\bm{e})$, so it maps any configuration
to its time-reversed partner at $-W$. Another relevant symmetry is \textit{spatial inversion $I$ (order 2).}
Reflection through the center of the ring reverses the cyclic ordering
$k \mapsto n{-}k \bmod n$ and simultaneously reverses each bond direction:
\begin{equation}
  I:\;(e_0,e_1,\ldots,e_{n-1}) \;\mapsto\; (-e_{n-1},-e_{n-2},\ldots,-e_0).
  \label{eq:I}
\end{equation}
$I$ satisfies $W(I\bm{e}) = -W(\bm{e})$. Besides $T$ and $I$, we start from the most symmetrical geometrical operation, \textit{Cyclic rotation $R$ (order $n$).} A counterclockwise permutation on the bond labels:
\begin{equation}
  R:\;(e_0,e_1,\ldots,e_{n-1}) \;\mapsto\; (e_{n-1},e_0,\ldots,e_{n-2}).
  \label{eq:R}
\end{equation}
$R$ generates the cyclic group $\mathbb{Z}_n$ and satisfies $W(R\bm{e}) = W(\bm{e})$. Since the classification for the most symmetrical $R$ at $n=3$ is trivial, we first use this case to illustrate the basic idea, and the other symmetry operations for $n=3$ are discussed below to demonstrate the applicability of the theory. It is straightforward to see $|W(g\bm{e})| = |W(\bm{e})|,\:\:g\in G=\langle R,T,I\rangle$. Therefore $|W|$ is a $G$-invariant, configurations in the same orbit share the same $|W|$.

\textit{\textcolor{blue} {Orbit enumeration via Burnside's lemma}}---The number of distinct equivalent classes(orbits) of the non-reciprocal responses can be found by Burnside's lemma~\cite{Burnside1911},
\begin{equation}
  |\mathcal{C}/G| = \frac{1}{|G|}\sum_{g\in G}|\mathrm{Fix}(g)|,
  \label{eq:burnside}
\end{equation}
where $\mathcal{C}=\{+1,-1\}^n$ is the set of all $2^n$ configurations,
$G$ acts on $\mathcal{C}$ by permuting configurations via symmetry
operations, and $\mathcal{C}/G$ denotes the set of equivalent classes termed orbits. An orbit is the quotient
of $\mathcal{C}$ by the equivalence relation $\bm{e}\sim\bm{e}'$ iff
$\bm{e}'=g\,\bm{e}$ for some $g\in G$. Here $\mathrm{Fix}(g)=\{\bm{e}: g\bm{e}=\bm{e}\}$ is the set of
configurations fixed by $g$ and $|\text{Fix}(g)|$ is the number of fix points of $g$. Each orbit represents physically equivalent configurations, i.e.\
a distinct type of non-reciprocal response.

\textit{\textcolor{blue}{Classification of $n=3$ symmetrical ring---}}We demonstrate the classification using the minimum non-trivial example, $n=3$. In this scenario, $W \in \{-3,-1,+1,+3\}$, with $W = 0$ strictly absent. The $2^3 = 8$ configurations distribute across these sectors as described in Table.~\ref{tab:orbits}. Since $|W|$ is a $G$-invariant, different $|W|$ cannot be in the same orbit. Applying the group action of $G = \langle R, T, I\rangle$ to all 8 configurations
(with the fixed-point counts tabulated in Table~\ref{tab:burnside_RTI} of SM~\ref{app:n3_RTI}), 
Burnside's lemma yields two distinct orbits and four subsets ($\langle R\rangle$-orbits) related by time reversal symmetry (TRS), listed in Table~\ref{tab:orbits}.

\begin{table}[t]
\centering
\caption{$T$-partner structure of $G=\langle R,T,I\rangle$ for the $n=3$ ring.
All $2^3=8$ configurations are accounted for.
The $T$-partner column shows the orbit related by time-reversal~\eqref{eq:T}. The two orbits under $G=\langle R,T,I\rangle$ are $\{A,B\}$ and $\{C,D\}$.}
\label{tab:orbits}
\begin{ruledtabular}
\begin{tabular}{cccll}
 set & $W$ & Size & Members & $T$-partner \\
\hline
$\mathcal{A}$ & $+3$ & 1
  & $(+,+,+)$
  & $\mathcal{B}$ \\
$\mathcal{B}$ & $-3$ & 1
  & $(-,-,-)$
  & $\mathcal{A}$ \\
$\mathcal{C}$ & $+1$ & 3
  & $(+,+,-)$,\;$(-,+,+)$,\;$(+,-,+)$
  & $\mathcal{D}$ \\
$\mathcal{D}$ & $-1$ & 3
  & $(-,-,+)$,\;$(+,-,-)$,\;$(-,+,-)$
  & $\mathcal{C}$ \\
\end{tabular}
\end{ruledtabular}
\end{table}

The configurations under $G=\langle R,T,I\rangle$ contain four $\langle R\rangle$-orbits, $\mathcal{A},\mathcal{B},\mathcal{C}$ and $\mathcal{D}$~(i.e. they can not be mapped to each other using $R$ alone). Burnside's lemma on $G$ further groups them into two different orbits. The orbit $\{\mathcal{A},\mathcal{B}\}$ represents \emph{pure circulation} orbit:
all bonds aligned, either fully counterclockwise ($W = +3$) or fully clockwise
($W = -3$). 
The orbit $\{\mathcal{C},\mathcal{D}\}$ represent \emph{semi-circulation}:
one bond reversed against the dominant flow. The rotation $R$ cycles between the
three members of each $\langle R\rangle$-orbit, so each has size three.

\textit{\textcolor{blue} { Generation of non-reciprocity---}}Using the two orbits in $G=\langle R, T, I\rangle$, we now establish the central symmetry condition for non-reciprocal transport.

For $n=3$, $T$ and $I$ induce \emph{identical pairings} on the set of
$\langle R\rangle$-orbits.
Direct computation shows that both map
$\mathcal{A}\leftrightarrow\mathcal{B}$ and $\mathcal{C}\leftrightarrow\mathcal{D}$. Both operations map $W\to-W$ at the orbit level and are identical as
permutations of the four orbits, even though they act differently on
individual configurations within each orbit.
This functional equivalence is rooted in the fact that for $n=3$, $W$
uniquely labels each $\langle R\rangle$-orbit. There is exactly one $\langle R\rangle$-orbit
per $W$ value, so both $T$ and $I$ are forced to pair the same orbits.

Non-reciprocity requires lifting the degeneracies
$\mathcal{A}\leftrightarrow\mathcal{B}$ and $\mathcal{C}\leftrightarrow\mathcal{D}$.
Since both $T$ and $I$ independently protect these degeneracies, each
alone is sufficient to maintain them. Therefore, both must be broken
simultaneously to generate non-reciprocity. This is a generic statement, independent of any microscopic model.
However, we note that even after breaking $T$ and $I$, the rotation $R$
still enforces a 3-fold internal degeneracy within $\langle R\rangle$-orbits, $\mathcal{C}$ and
$\mathcal{D}$.
Therefore, even breaking $T$ and $I$ still does not generate the responses $\mathcal{C}$ and $\mathcal{D}$.  On the other hand, the responses associated with $\mathcal{A}$ and $\mathcal{B}$ ($W=\pm3$, circulator) are singletons under $R$ and are accessible as soon as $T$ and $I$ are broken.  The justification of this argument using our toy model is present in SM~\ref{app:Geo}.

\textit{\textcolor{blue} {Isosceles geometry---}}When the geometry is made \emph{isosceles} (one bond distinguished from
the other two), the 3-fold rotation $R$ is broken.
The surviving symmetry is the mirror-time symmetry $\sigma=IT$ (order~2) that can remain when $R$ is broken. It swaps the two equal legs without sign flip:
\begin{equation}
  \sigma = IT:\;(e_0,e_1,e_2)\mapsto(e_2,e_1,e_0),\quad W\mapsto W.
  \label{eq:sigma_def}
\end{equation}
 The full symmetry group reduces to
\begin{equation}
  G_\mathrm{iso} = \langle T,I,\sigma\rangle = \langle T,I\rangle
  \cong \mathbb{Z}_2\times\mathbb{Z}_2,\quad |G_\mathrm{iso}|=4,
  \label{eq:group_iso}
\end{equation}
since $\sigma=IT\in\langle T,I\rangle$ introduces no new generator.
Applying Burnside's lemma to $G_\mathrm{iso}$
(Table~\ref{tab:burnside_iso} in SM~\ref{app:iso}) yields three
symmetry classes, listed in Table~\ref{tab:orbits_iso}.

\begin{table}[t]
\centering
\caption{Orbit classification for the isosceles $n=3$ ring under
$G_\mathrm{iso}=\langle T,I\rangle\cong\mathbb{Z}_2\times\mathbb{Z}_2$.
The defect (minority bond current) position distinguishes orbits~II and~III,
both at $|W|=1$.
All $2^3=8$ configurations are accounted for: $2+2+4=8$.}
\label{tab:orbits_iso}
\begin{ruledtabular}
\begin{tabular}{ccccc}
Orbit & $W$ & Size & $\mathrm{Stab}$ & Members \\
\hline
I   & $\pm3$ & 2 & $\{e,\sigma\}$
  & $(+,+,+)$,\;$(-,-,-)$ \\
II  & $\pm1$ & 2 & $\{e,\sigma\}$
  & $(+,-,+)$,\;$(-,+,-)$ \\
III & $\pm1$ & 4 & $\{e\}$
  & $(+,+,-)$,\;$(-,+,+)$,\;$(+,-,-)$,\;$(-,-,+)$ \\
\end{tabular}
\end{ruledtabular}
\end{table}

The three classes are characterized by two quantities: (i) the winding
magnitude $|W|$ and (ii) the \emph{defect position}, whether the minority bond
sits on the base ($e_1$, orbit~II) or on a leg ($e_0$ or $e_2$, orbit~III).
Orbit~I and~II each has non-trivial stabilizer $\{e,\sigma\}$, $\sigma$
fixes their representatives because swapping the legs leaves them unchanged.
Orbit~III has trivial stabilizer $\{e\}$. $\sigma$ can not fix it since $\sigma$ moves the defect from one leg to the other.

When $T$ and $I$ are broken and $\sigma$ is preserved, the degeneracies
within orbits~I and~II are lifted(both are $T$-doublets of size~2), so these classes generate observable non-reciprocal responses, $W=\pm3$ (circulator) and $W=\pm1$ on the base bond (semi-circulator).
Orbit~III, however, remains forbidden since its four-member orbit contains
two pairs related by $\sigma$
, so the internal
degeneracy is not fully lifted.
We therefore predict, for the isosceles $n=3$ ring, breaking $T$ and $I$ generates only
orbit~I and orbit~II non-reciprocal responses. The semi-circulator response only allows flow along the minority bond. Orbit~III responses are
symmetry-forbidden as long as $\sigma$ survives. This prediction is independent of the microscopic detail and  validated in a toy model.%

\begin{figure}[b]
\includegraphics[width=0.25\textwidth]{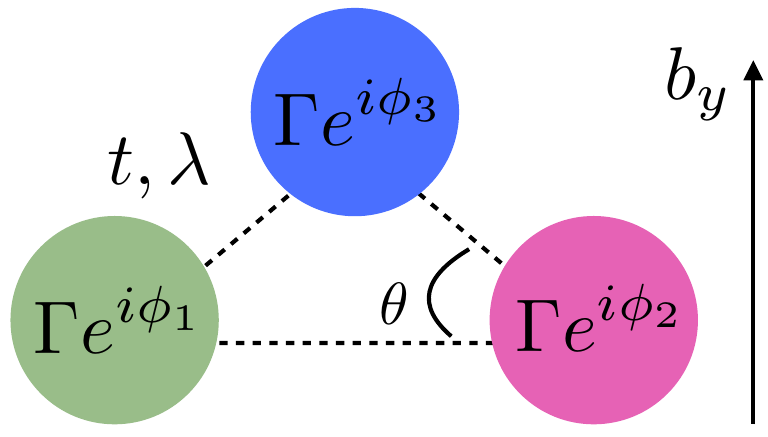}
\caption{ Minimal phenomenological three-site model to describe the generation of non-reciprocal circulation in a triple quantum dot triangle. The cyclic rotation $R$-symmetry is broken to the reflection  when $\theta\neq\pi/3$ or $b_y\neq0$.   }\label{fig:sys}
\end{figure}
\textit{\textcolor{blue} {Toy Model for $n=3$: Poor man's Majorana Triangle---}}We now validate the group-theoretic predictions using a minimal model. The two predictions to test are: (i) non-reciprocity requires simultaneous
breaking of both $T$ and $I$, and (ii) only orbit~I and orbit~II responses
appear in the isosceles geometry where $\sigma$ is preserved.

\begin{figure*}[t]
\includegraphics[width=0.95\linewidth]{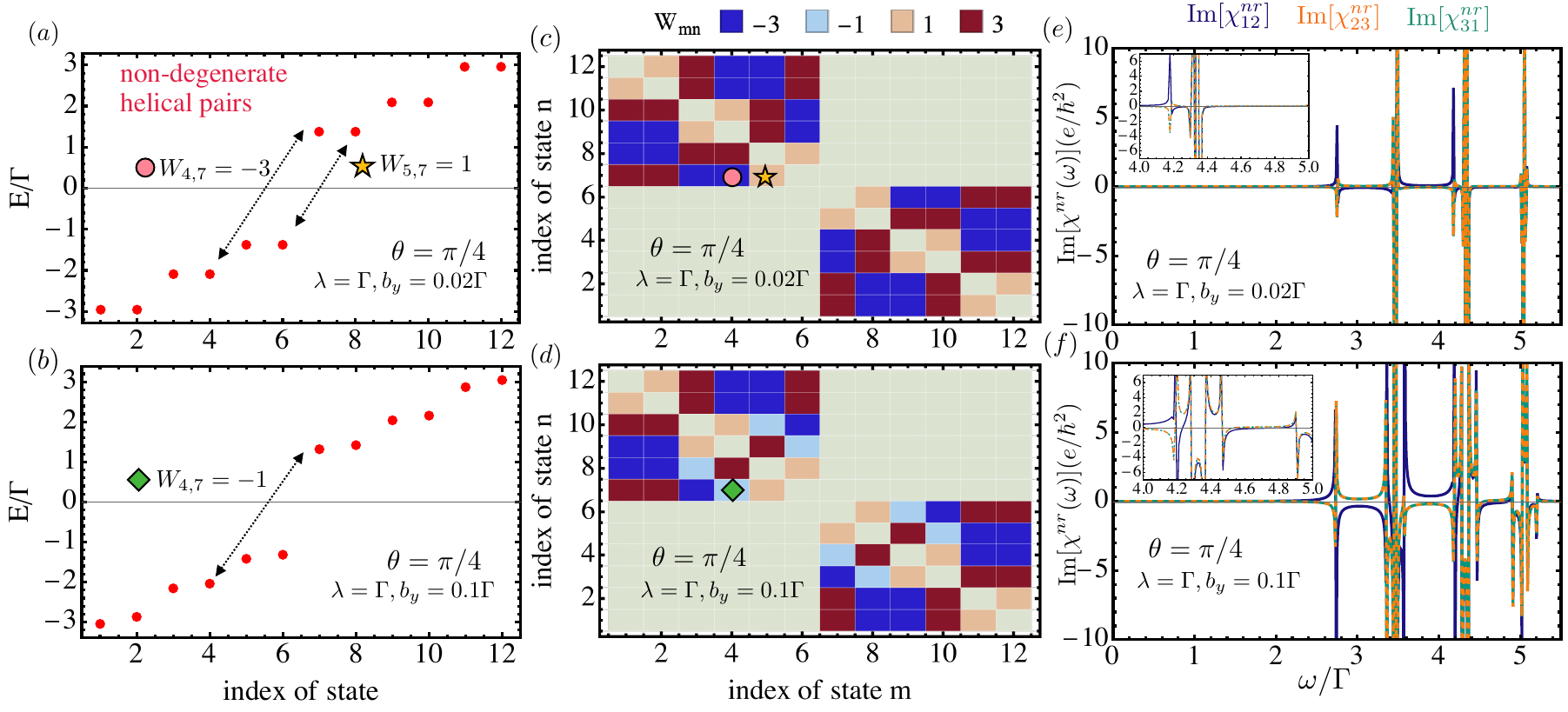}
\caption{(a,b) Single particle energy spectrum for the minimal model at $\theta=\pi/4$ for (a) $b_y=0.02\Gamma$ (b) $b_y=0.1\Gamma$. We also indicate the value of the $W_{mn}$ for some of the intra(inter)-helcial pair transitions by the black arrows and labels in (c,d). (c,d)  $W_{mn}=\sum_{a>b}\text{sign}[ \text{Im}(J_{a;mn} J_{b;nm})]   [ n_f(\epsilon_n)-n_f(\epsilon_m)]$ for the minimal model to characterize the direction of (semi-) circulator response for different excitations. (e,f) Non-reciprocal reactive response Im$[\chi^{nr}]$ for the devices for (e) $b_y=0.02\Gamma$ and  (f) $b_y=0.1\Gamma$. The  inset shows the  circulator response band as we enhance the splitting by enhancing $b_y$ from (e) to (f).
}\label{fig:chi}\label{fig:cir}\label{fig:Rs}
\end{figure*}

We consider three quantum dots connecting to superconducting baths and the hopping between them is spin dependent given by Rashba spin-orbital coupling (SOC). The qualitative behavior can be described in the atomic limit where the superconducting baths are integrated out assuming the SC gap $\Delta_0\to\infty$~\cite{AL_Meden_2019,QD_GAL_PhysRevB.107.115407}. Denoting the spinor of the three sites, $\Psi=(\Psi_1,\Psi_2,\Psi_3)$ and $\Psi_a = (d_{\uparrow a},d_{\downarrow a},d^\dag_{\uparrow a},-d^\dag_{\downarrow a})$ with $\{d_{\sigma a},d_{\sigma'b}^\dag\}=\delta_{ab}\delta_{\sigma\sigma'}$ the fermionic operators, the Hamiltonian is parametrized by,
\begin{align}\label{eq:Hmin}
    \hat{H} & = \begin{pmatrix}
    H_1&T_{12}&T_{13}\\
    T_{21}&H_2&T_{23}\\
    T_{31}&T_{32}&H_3
    \end{pmatrix},
\end{align}
where
\begin{align}
    H_a&=  \tau_z \sigma_y b_y +\Gamma\left(\tau_x \cos\phi_a-\tau_y \sin\phi_a\right)\sigma_x,\\
    T_{ab}&= -t \sigma_0\tau_z+i\lambda\left( \sigma_x\tau_z\sin\theta_{ab}-\sigma_y\tau_0\cos\theta_{ab}\right),\\
    T_{ba}&=T_{ab}^\dag.
\end{align}
The minimal atomic limit model describes three sites with onsite induced pairing potential $\Gamma e^{i\phi_a}$ and Zeeman field $b_y$. The spin-orbital coupling is parametrized by a spin-dependent hopping with $t$ and $\lambda$. 
$\theta_{ab}$ are defined by the angles of the three edges of the triangle area in Fig.~\ref{fig:sys} (a). We parametrize them using $\theta_{12}=0, \theta_{23}=\pi-\theta$ and $\theta_{31}=\pi+\theta$ to capture the orbital effect of SOC. 
The $R$-symmetry is broken by finite $b_y$ or $\theta\neq\pi/3$, and the responses are described by the classification on $G_\text{iso}$.

Using $T=iI_3\sigma_y\tau_0\mathcal{K}$, we find
$T H(b_y,\phi_a) T^{-1}=H(-b_y,\pi-\phi_a)$.
Both the Zeeman field $b_y$ and the SC phase biases break TRS. On the other hand, the inversion operator is $I_{ab}=\mathcal{S}_{ab}\sigma_x\tau_0$,
where $S_{ab}$ is the permutation matrix. We find $I_{ab}HI_{ab}^{-1}\neq H$ unless $\lambda=0$
and $\phi_a=\phi_b$ for some pair of $a\neq b$. Therefore, IS is broken by SOC and SC phase biases. 
Both $T$ and $I$ must be broken simultaneously to generate the non-reciprocity. In this model, this is achieved by (i) finite SC phase biases, or (ii) in-plane Zeeman field $b_y$ + SOC $\lambda$. The former has been discussed elsewhere~\cite{Pauli_PhysRevLett.132.046002}. We focus on the latter. Finally, we note $\sigma_{12}=I_{12}T$ is preserved in this case.

We analyze the non-reciprocal AC response within the linear response framework. 
The non-reciprocal linear response is described by the current-current response function, $\chi^{nr}_{ab}(\omega)=  \chi_{ab}(\omega)-\chi_{ba}(\omega)$ where $\chi_{ab}(\omega)$ is given by the Kubo formula for single particle eigen-states~\cite{Kubo_doi:10.1143/JPSJ.12.570},
\begin{align}\label{eq:chi}
\chi_{ab}(\omega) &=   \sum_{mn}  \frac{n_f(E_{n}) - n_f(E_{m})}{\hbar \omega - (E_{m} - E_{n}) + i\delta}  J_{a;nm}  J_{b;mn}  .
\end{align}
The current operator is defined as $\hat{J}_a=- \partial \hat{H}/\partial \phi_a=\Gamma(\tau_x\sin\phi_a+\tau_y\cos\phi_a)\sigma_x$ and $T\hat{J}_aT^{-1}=-\hat{J}_a$ if TRS is preserved in $\hat{H}$. $J_{a;mn}=\langle m|\hat{J}_a|n \rangle$ is the matrix element and $n_f(\epsilon)$ is the Fermi function.


From Eq.~\eqref{eq:chi}, the reactive response is given by $\text{Im}[\chi^{nr}_{ab}] =   \sum_{mn}  P\left(\frac{n_f(E_{n}) - n_f(E_{m})}{\hbar \omega - (E_{m} - E_{n})  } \right) \text{Im}[J_{a;nm}  J_{b;mn}]   $ where $P$ denotes the principal value. The expression shows several discontinuities near the resonances at $\hbar \omega = E_{m} - E_{n}$ [c.f. Fig~\ref{fig:chi} (e,f)]. The circulator behavior $(1\to2\to3$ or vice versa) can be characterized by the sign of $\text{Im}[\chi^{nr}]$ near these discontinuities which is described by $W_{mn} = \sum_{a>b}\text{sign}[ \text{Im}(J_{a;mn} J_{b;nm})]   [ n_f(\epsilon_n)-n_f(\epsilon_m)]$ and $W_{mn}=\pm3$ for a circulator response.

Under zero field, the system shows degenerate helical Andreev bound state (ABS) pairs due to the TRS. The degeneracy is lifted by breaking the TRS with a Zeeman field. In Fig.~\ref{fig:cir} (a,b) we show the single particle spectrum for the minimal model with finite SOC under weak to intermediate in-plane Zeeman field. The ABS features split helical pairs. 

In Fig.~\ref{fig:cir} (c) we demonstrate the $W_{mn}$ factor of the system under $b_y=0.02\Gamma$. At weak field, we only find circulator response $(W_{mn}=\pm3)$ realized for the inter-helical pair excitations  and the semi-circulator responses for intra-helical pair excitations. On the contrary, in panel (d) we demonstrate the situation under a stronger in-plane Zeeman-field along the reflection axis.  Semi-circulator response ($W=\pm1$) are generated also for inter-pair excitations. This is because the strong field mixes the originally orthogonal helical pair eigen-states that changes the overlap matrix elements for the responses. Finally we note only $W_{mn}=\pm1$ and $\pm3$ are observed in the responses as predicted for the isosceles geometry.

In Fig.~\ref{fig:Rs}(e,f), we demonstrate the reactive responses of the system with these two different $b_y$. As expected, the reactive non-reciprocal responses only have  two different types. Either all the discontinuities are in the same signs (circulator, $W=\pm3$) or one of them ($\text{Im}[\chi^{nr}_{12}]$) is different from the rest (semi-circulator, $W=\pm1$). The latter is exactly the orbit~II response predicted in our theory. Specifically, from the inserts of panel (e) to (f) we observe the circulatory response peaks further split and result in a circulator band.  This is because the splitting of the helical pair energy levels are enhanced by the larger Zeeman field.  Furthermore, we note the formation of the circulatory band is between the absorption peaks and fully reactive~\cite{reactiveR_PhysRevB.38.9581,Pauli_PhysRevLett.132.046002}.

\begin{figure}[t]
\includegraphics[width=0.75\columnwidth]{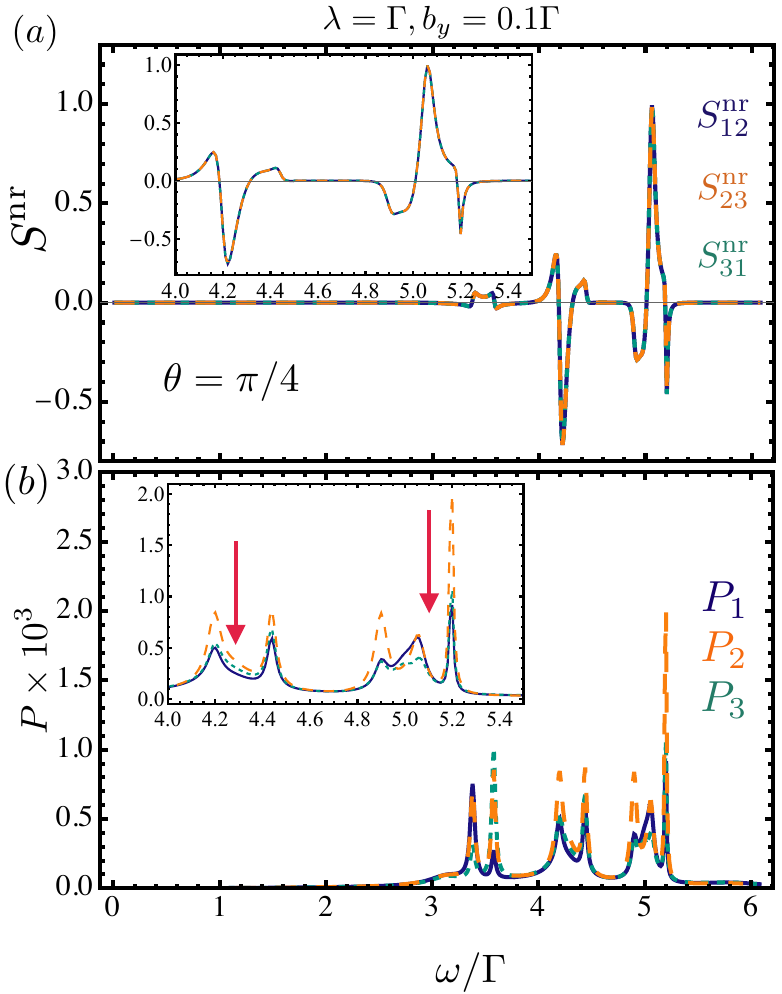}
\caption{ Characterization of the system with in-plane field. (a) is the non-reciprocal transmission of the device. The red arrows indicate the  reactive response region of the device.
(b) show the absorption spectra of the three terminals of the device.  }\label{fig:S} 
\end{figure}
\textit{\textcolor{blue} {Reactive AC microwave circulator---}}The emergent circulatory band in Fig.~\ref{fig:Rs}(f) means the possibility of reactive response in a certain frequency region. To quantify the working behavior of the toy model, we calculate the AC scattering parameters using the following formula for the microwave transmission scattering matrix~\cite{microwave_Collin2001},
\begin{align}
S= (I_3 - \sqrt{Z} Y\sqrt{Z})( I_3 + \sqrt{Z} Y\sqrt{Z} )^{-1},
\end{align}
where $Z = \text{diag}(z_1,z_2,z_3)=(6.4,6.4,4.4) \hbar^2/e$ is the impedance matrix with diagonal elements from the impedances of the SC terminals, and $Y_{ab}(\omega) =  i\chi_{ab}(\omega)/\omega$ is the admittance matrix. The impedence of order $\hbar^2/e$ is a natural choice of the quantum dot model~\cite{imp_BUTTIKER1993364}.  To characterize the device, we define the non-reciprocal transmission between terminal $a$ and $b$, $ S_{ab}^{nr}  = |S_{ab}|^2- |S_{ba}|^2$. 
The circulator is characterized by the condition  that all $S^{nr}_{a>b}$ have the same signs and a perfect circulator behavior is indicated by $|S^{nr}_{a>b}|=1$. We define the absorption of microwave power, $P_{a;\text{lost}}= 1- \sum_{b} |S_{ba}|^2$ to measure the power lost during the transmission.
Fig.~\ref{fig:S} (a) is the non-reciprocal transmission of the toy model. Several circulator response bands are observed. Panel (b) is the power absorption of the system. In the inset we indicate the reactive response region where dips appear (red arrows). We note the absorption is small in the limit of tiny broadening $\delta\to0^+$ of Eq.~\eqref{eq:chi}. In SM~\ref{app:bd}, we also include the effect of finite broadening $\delta=0.001\Delta$ for comparison. The reactive feature remains observable.

\textit{\textcolor{blue} {Discussion and conclusion---}}
In conclusion, we have described a generic mathematical framework to classify the non-reciprocal response of $n$-terminal ring device.  Including the geometrical symmetries along with the Burnside's lemma can efficiently predict the direction of non-reciprocal responses. Extension on the theory with $n>3$ along with different geometrical symmetries can give hints on the design of non-reciprocal $n$-terminal devices for future studies.  A more complicated abstract construction beyond ring graph and analysis incorporating graph theory will be investigated elsewhere.  Finally, this theory requires the systems near the high-symmetry points. If there is no symmetry remaining after breaking $T$ and $I$, it is pointless to apply the group classification.

Besides the mathematical framework, we also predict a reactive circulator response can be achieved using the poor man's majorana triangle system with broken inversion and time reversal symmetries.  

\textit{\textcolor{blue}{Acknowledgments--}}C.-H.H has been supported by EU's HORIZON-RIA Programme under Grant No. 101135240 (JOGATE). This work is part of the Finnish Centre of Excellence in Quantum Materials (QMAT). 

\nocite{apsrev42Control}
\bibliographystyle{apsrev4-2} 
\bibliography{ref_G}



\clearpage
\newpage

\onecolumngrid
\begin{center}
    \textbf{\large Supplemental Material for Symmetry Classification of Non-Reciprocal Responses in Multiterminal Ring Devices} \\
    \vspace{10pt}
\end{center}
\setcounter{section}{0}

\section{Prediction of non-reciprocal responses with geometrical symmetry operations}\label{app:Geo}
\begin{figure}[h]
\includegraphics[width=0.8\columnwidth]{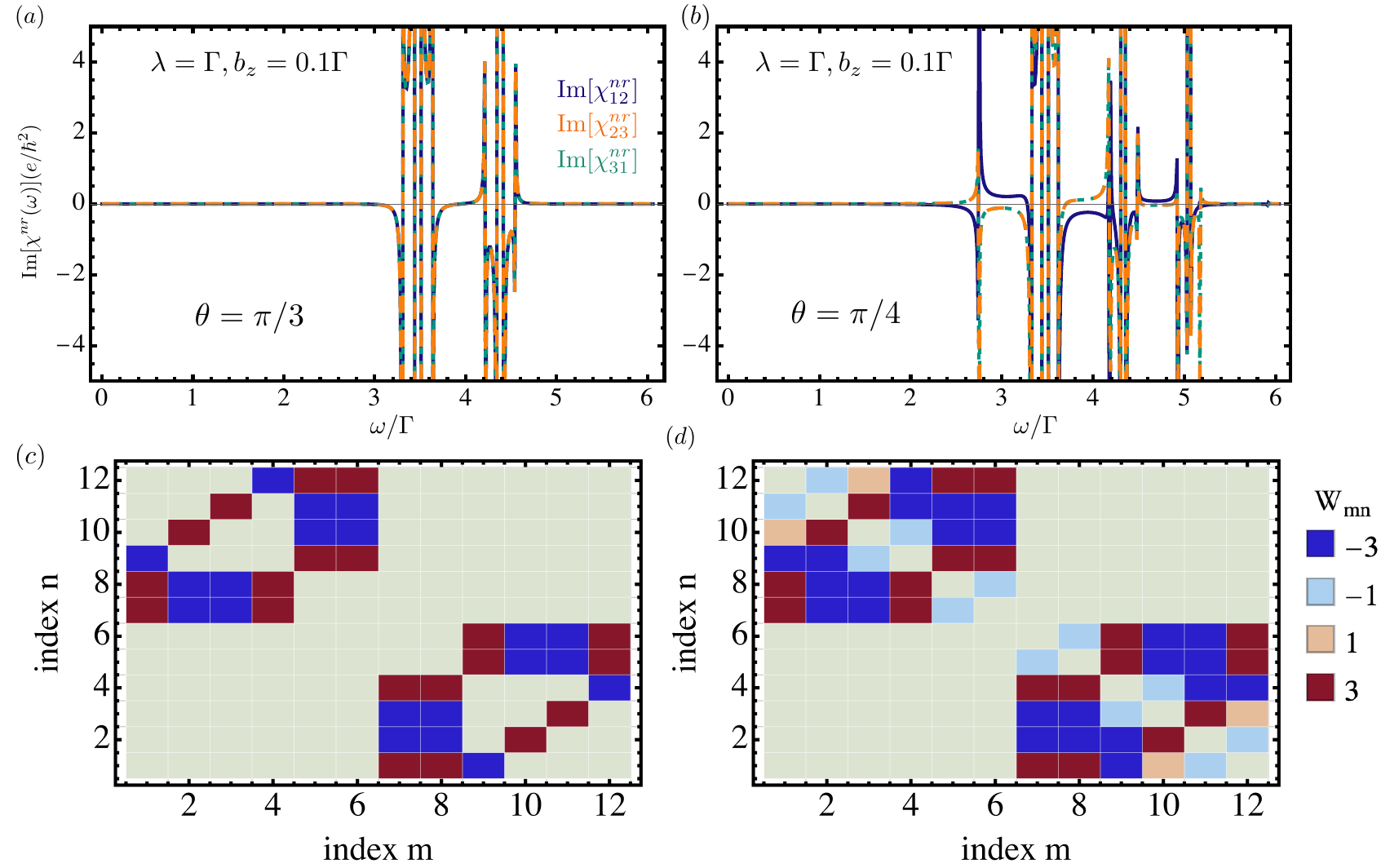} 
\caption{Reactive part of the response function $\text{Im}[\chi^{nr}]$ for (a) $\theta=\pi/3$ (b) $\theta=\pi/4$. (c,d) $W_{mn}=\sum_{a>b}\text{sign}[ \text{Im}(J_{a;mn} J_{b;nm})]   [ n_f(\epsilon_n)-n_f(\epsilon_m)]$ for (a,b).}\label{fig:bz}
\end{figure}
In this section, we supplement the prediction made in the main text for the system with different geometrical symmetry operations. We consider the Zeeman field in the out-of-plane z-direction, therefore it does not breaks the cyclic rotation $R$. $R$ is broken by tuning the system geometry angle $\theta$.
In Fig.~\ref{fig:bz} (a), $\theta=\pi/3$, the system preserves $R$. Therefore, breaking $T$ and $I$ only generates non-reciprocal responses of class $\{\mathcal{A},\mathcal{B}\}$ in Tab.~\ref{tab:orbits} namely all the $\text{Im}[\chi^{nr}_{ab}]$ have the same sign. This is also indicated in (c) where all the $|W_{mn}|=3$. On the contrary, in panel (b) we pick $\theta=\pi/4$. The classification predicts the responses of classes I and II in Tab.~\ref{tab:orbits_iso} which means either all the responses have the same signs($|W_{mn}|=3$) or $\text{Im}[\chi_{12}^{nr}]$ is in a different sign to the rest($|W_{mn}|=1$). Both of the figures are consistent with our prediction based on the group classifications.
\clearpage
\newpage
\section{Effect of finite broadening}\label{app:bd} 
\begin{figure}[h]
\includegraphics[width=1\columnwidth]{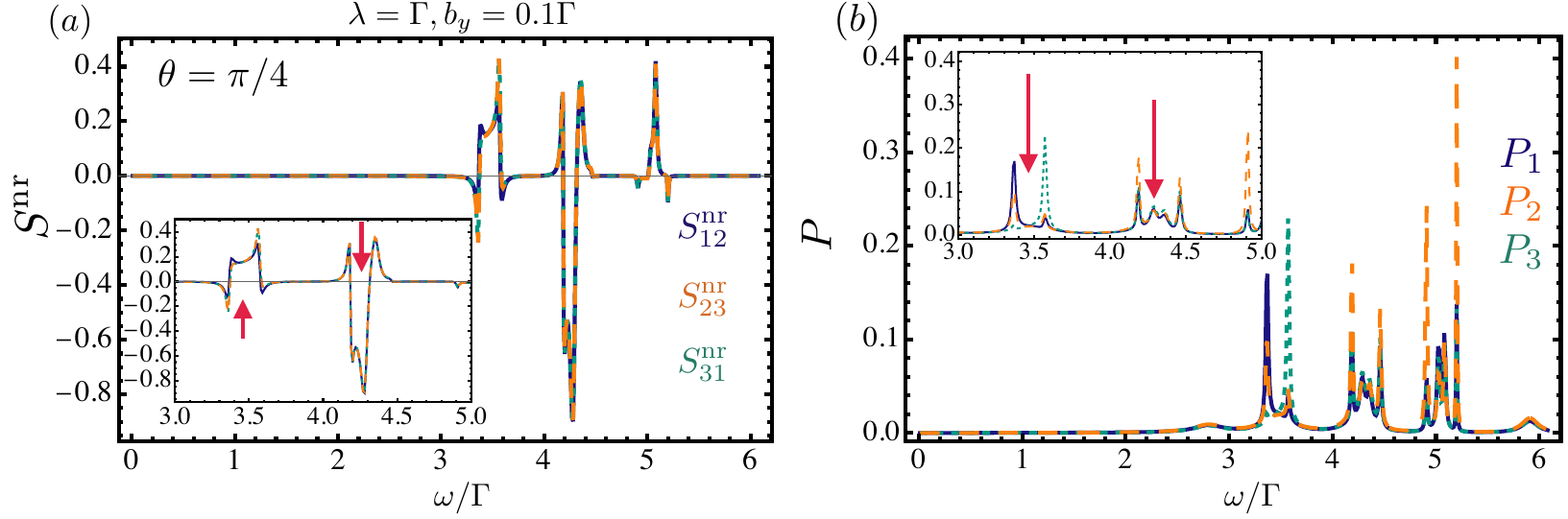} \label{fig:Sbd}
\caption{(a) Non-reciprocal transmission and (b) power absorption of the toy model including the finite broadening and temperature $k_BT=\delta=0.001\Gamma$.}
\end{figure}

In this appendix, we consider the effect of broadening on the device's behavior. The effect of broadening appears in Eq.~\eqref{eq:chi} in two ways. The Fermi-Dirac distribution acquires the temperature dependence, and the broadening parameter is no more infinitesimal but $\delta\simeq k_BT$. Here we pick $k_BT = 0.001\Delta$. Including both effect, we find the absorption spectrum and non-reciprocal transmission in Fig.~\ref{fig:Sbd}. The thermal effect broadens the absorption peaks and leads to an overall enhancement on the absorption. However, the reactive response region remains.  In these plots, we pick $(z_1,z_2,z_3)=(0.78,0.78,0.9)\hbar^2/e$.

\section{Splitting of ABS energies by inplane field and SOC}\label{app:eabs} 
\begin{figure}[h]
\includegraphics[width=1\columnwidth]{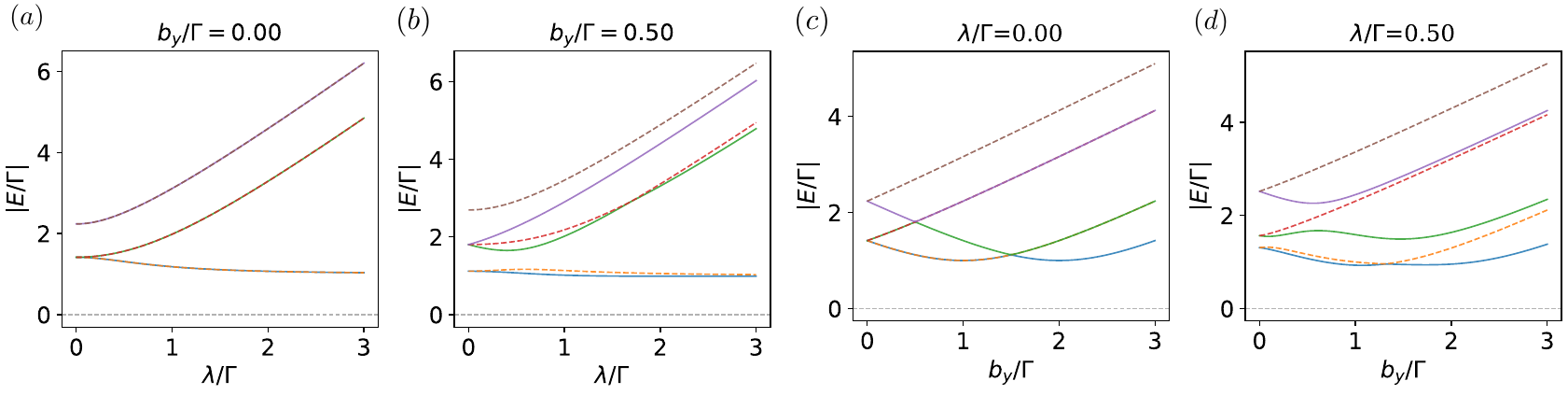} \label{fig:eabss}
\caption{(a,b) ABS energies as a function of $\lambda$ under two different $b_y$  (c,d) ABS energies as a function of $b_y$ under two different $\lambda$. In all the panels, $t=\Gamma=1$ and $\Theta=\pi/3$.}
\end{figure}

In this section, we discuss the effect of inplane Zeeman field and SOC on the ABS energies of the toy model.  At zero $\lambda$ and $b_y$, ABS energy can be obtained by discrete Fourier transformations as $|E_4|=\sqrt{\Gamma^2+t^2}$ (4 fold degeneracy) and $|E_2|=\sqrt{\Gamma^2+4t^2}$ (2 fold degeneracy). The former corresponds to the discrete momentums $k=\pm2\pi/3$ and the latter corresponds to $k=0$. The degeneracies are doubled by including the spin components. This is illustrated in Fig.~\ref{fig:eabss} (a)(c) the energy at zero $\lambda$ and $b_y$. 
The SOC $\lambda$ transforms the spin degenerate ABS energies into helical ABS pairs which further split the degenerate energies of the $k=\pm2\pi/3$ pair leading to three doubly degenerated pairs. This is indicated in (a). On the contrary, the effect of $b_y$ is to split the spin degeneracies leading to a splitting on all the degenerate spin pairs as shown in panel (c). When both terms are non-zero, the magnitude of the ABS energy split into six different values in panels (b)(d).

\section{Topological phases of the minimal model }\label{app:A} 

\begin{figure}[h]
    \centering    \includegraphics[width=1\textwidth]{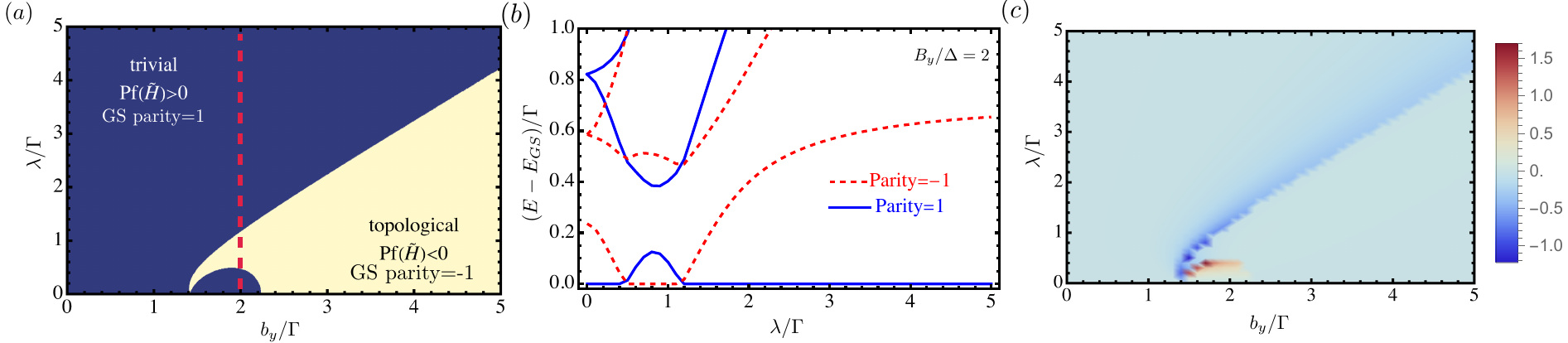}
    \caption{(a) Topological phase diagram of the minimal model with $\theta=\pi/4,\:t=\Delta$ and $\phi_a=0,\forall a$. (b) Low-lying many-body energy levels relative to the ground state energy along the red dashed path in panel (a). (c) Overall chiral DC conductance, $G^{nr}=G_{12}^{nr}+G_{23}^{nr}+G_{31}^{nr}$ (in unit of $e^2/\hbar$) for the system. }
    \label{fig:app1}
\end{figure}


We note there is a non-trivial topological phase transition in the minimal model. The phase boundary can be found by locating the emergence of the zero ABS.  And the phases can be characterized by the Pfaffian of the Hamiltonian in the Majorana basis, $ a_{1\sigma a} = (d^+_{\sigma a}+d_{\sigma a})/\sqrt{2},\: a_{2\sigma a} = i(d^+_{\sigma a}-d_{\sigma a})/\sqrt{2}$. Denoting the Hamiltonian in the Majorana basis as $\tilde{H}$, we have $\tilde{H} =  U HU^\dag$ where $U = I_3 u/\sqrt{2}$ and,
\begin{equation}
u =
\begin{pmatrix}
    1&0&1&0\\
    0&1&0&-1\\
    -i&0&i&0\\
    0&-i&0&-i
\end{pmatrix}.
\end{equation}
The phase diagram is characterized by the Kitaev-Akhmerov invariant, $Q=\text{sign}[\text{Pf}(\tilde{H})]$~\cite{A_Yu_Kitaev_2001}. A topologically  non-trivial phase~($Q<0$) is obtained at strong zeeman field. In panel (b) we show the many-body energy levels along the red-dashed path in panel (a). Level crossings are observed at the transition points. The ground states have different parities across the phase boundary. 

We also provide the chiral DC conductance $G_{ab}^{nr} = G_{ab}-G_{ba}$, $G_{ab}=\lim_{\omega\to0} i\chi_{ab}(\omega)/\omega$. Unlike the AC case, there is a strong enhancement of the chiral conductance near the transition.

\section{The absolute non-reciprocal weight}
\begin{figure}[h]
\includegraphics[width=\columnwidth]{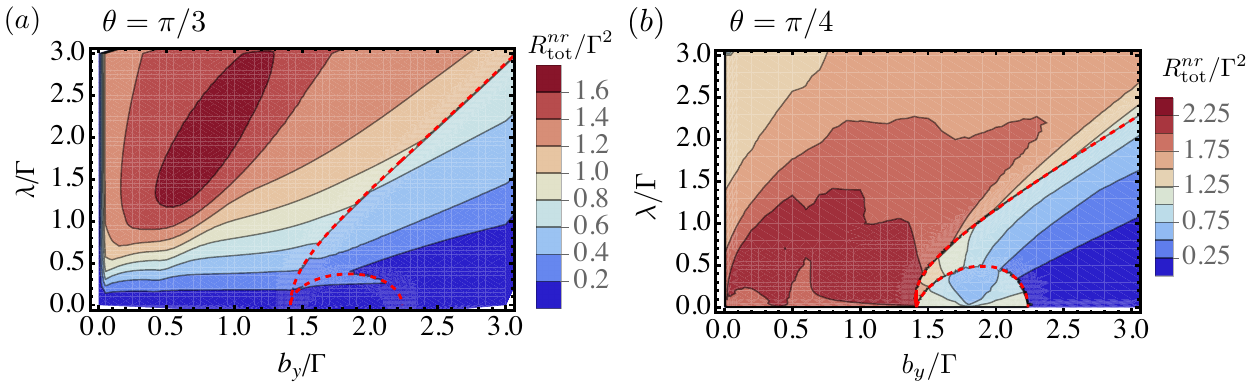}
\caption{   Absolute non-reciprocal  weights of the toy model for (a) $\theta=\pi/3$ and (b) $\theta=\pi/4$ under SOC+in-plane field. The non-reciprocity is characterized by non-zero $R^{nr}_\text{tot}$. The red dashed line indicates the emergence of zero energy ABS in the toy model. In both figures we pick $t=\Delta$. } \label{fig:Rs2}
\end{figure}

The non-reciprocity originates from the non-zero value of $\text{Im}[\langle n | \hat{J}_a | m \rangle \langle m | \hat{J}_b | n \rangle]$, we introduce the absolute non-reciprocal weight, $R^{nr}_\text{tot}=\sum_{a>b} R_{ab}^{nr}$, that sums all the absolute values of the non-reciprocal weights. The absolute non-reciprocal weight acts as a metric to probe the non-reciprocity induced by ABS at finite frequencies:
\begin{align}
R^{nr}_{ab} &=
\int_0^{2\Delta_0} \biggl|
\sum_{mn} \text{Im}\left[\langle n | \hat{J}_a | m \rangle \langle m | \hat{J}_b | n \rangle\right] \left[ n_f(E_{n}) - n_f(E_{m}) \right] \delta(E- E_m+E_n )  
\biggr| dE, 
\end{align}
which sums all the absolute values of the non-reciprocal weights at at transition energies $0<E<2\Delta_0$ inside the BCS gap. In the effective atomic-limit toy model, $\Delta_0\to\infty$.
In Fig.~\ref{fig:Rs} (c,d), we plot the absolute weights $R_\text{tot}^{nr}=\sum_{a<b}R_{ab}^{nr}$. In Fig.~\ref{fig:Rs2} (c,d), we plot the absolute weights $R_\text{tot}^{nr}$ to understand the effect of $\lambda$ and $b_y$ on the overall non-reciprocity at different $\theta$. The most obvious feature is given by the red dashed boundary which corresponds to the topological phase transition where the ground state changes it parity (see SM~\ref{app:A}). However, the non-reciprocity at this phase is much smaller than the trivial phase. In the trivial phase, there exists an optimal region which emerges at finite $\lambda$ and $b_y<1.5\Gamma$. Furthermore, we note the optimal region at $\theta=\pi/4$ requires a smaller $\lambda$ and demonstrates a larger $R_\text{tot}^{nr}$ than the case of $\theta=\pi/3$.
 \section{Classification of the $n=3$ ring under $G=\langle R,T,I\rangle$}
\label{app:n3_RTI}

The equilateral $n=3$ ring has full symmetry group
$G=\langle R,T,I\rangle$ of order~12.
Applying Burnside's lemma~\eqref{eq:burnside} with the fixed-point counts
in Table~\ref{tab:burnside_RTI} gives $24/12=2$ symmetry classes.

\begin{table}[h]
\centering
\caption{Fixed-point counts for $G=\langle R,T,I\rangle$ ($|G|=12$)
acting on the $n=3$ configuration space.
The TRS and inversion families each contribute zero
(both require $e_i=-e_i$, impossible for $e_i=\pm1$).
The $IT$-family contributes 12 because $IT$ is a sign-free permutation
($IT:\,(e_0,e_1,e_2)\mapsto(e_2,e_1,e_0)$), fixing any configuration
with $e_0=e_2$.}
\label{tab:burnside_RTI}
\begin{ruledtabular}
\begin{tabular}{llr}
Element $g$ & Fixed-point condition & $|\mathrm{Fix}(g)|$ \\
\hline
$e$    & all configurations & 8 \\
$R$    & $e_0=e_1=e_2$ & 2 \\
$R^2$  & $e_0=e_1=e_2$ & 2 \\
\hline
$T$, $TR$, $TR^2$   & $e_i=-e_i$: impossible & 0 \\
\hline
$I$, $IR$, $IR^2$   & leads to $e_i=-e_i$: impossible & 0 \\
\hline
$IT$    & $e_0=e_2$: $(+,+,+),(+,-,+),(-,+,-),(-,-,-)$ & 4 \\
$ITR$   & $e_0=e_1$: $(+,+,+),(+,+,-),(-,-,+),(-,-,-)$ & 4 \\
$ITR^2$ & $e_1=e_2$: $(+,+,+),(+,-,-),(-,+,+),(-,-,-)$ & 4 \\
\hline
\multicolumn{2}{r}{Burnside: $24/12$} & $=\mathbf{2}$ \\
\end{tabular}
\end{ruledtabular}
\end{table}

The two classes each contain two $\langle R\rangle$-orbits connected by
$T$ and $I$ (Table~\ref{tab:orbits}).
Both $T$ and $I$ map $\mathcal{A}\leftrightarrow\mathcal{B}$ and
$\mathcal{C}\leftrightarrow\mathcal{D}$, so the orbit pairs
$\{\mathcal{A},\mathcal{B}\}$ and $\{\mathcal{C},\mathcal{D}\}$ each
form one symmetry class(orbit).

\section{Classification of isosceles Configurations: $n=3$}
\label{app:iso}


The equilateral triangle possesses the full dihedral symmetry group
$D_3 = \langle R, I \rangle$, where $R$ is the $2\pi/3$ rotation and
$I$ is the spatial inversion (swap of vertices $1\leftrightarrow 2$).
When the triangle is made \emph{isosceles} where one bond (the base $e_1$)
distinguished from the two equal legs ($e_0$, $e_2$), the three-fold
rotation $R$ is broken while the reflection through the apex axis survives.
We analyze how this geometric symmetry reduction affects the orbit
classification of current configurations.

\subsection{Symmetry group of the isosceles ring}

The isosceles $n=3$ ring retains three symmetry operations:

\paragraph{Time-reversal $T$ (order 2).}
\begin{equation}
  T:\;(e_0,e_1,e_2)\mapsto(-e_0,-e_1,-e_2),\quad W\mapsto -W.
\end{equation}

\paragraph{Spatial inversion $I$ (order 2).}
Reflection through the apex axis fixes vertex~0 and swaps
vertices $1\leftrightarrow 2$, reversing all bond orientations:
\begin{equation}
  I:\;(e_0,e_1,e_2)\mapsto(-e_2,-e_1,-e_0),\quad W\mapsto -W.
\end{equation}
\paragraph{Leg-swap $\sigma$ (order 2).} This is the unique reflection symmetry of the isosceles geometry.
The rotation $R$ is absent since the base bond $e_1$ is inequivalent to
the leg bonds $e_0,e_2$, breaking the three-fold symmetry. The product $\sigma = IT$ acts as a sign-free permutation of the
two leg bonds:
\begin{equation}
  \sigma = IT:\;(e_0,e_1,e_2)\mapsto(e_2,e_1,e_0),\quad W\mapsto W.
  \label{eq:sigma_def}
\end{equation}
Since $\sigma = IT$, it is already contained in $\langle T,I\rangle$
and introduces no new generator.
The full symmetry group is therefore
\begin{equation}
  G_\mathrm{iso} = \langle T,\,I,\,\sigma\rangle
  = \langle T,\,I\rangle
  \cong \mathbb{Z}_2\times\mathbb{Z}_2,
  \qquad |G_\mathrm{iso}| = 4,
  \label{eq:group_iso}
\end{equation}
the Klein four-group with elements $\{e,\,T,\,I,\,\sigma\}$.

\subsection{Burnside enumeration and the three symmetry classes}

Applying Burnside's lemma~\eqref{eq:burnside} to $G_\mathrm{iso}$
acting on $\mathcal{C} = \{+1,-1\}^3$:

\begin{table}[h]
\centering
\caption{Fixed-point counts for the isosceles $n=3$ ring.
Neither $T$ nor $I$ fixes any configuration (both require $e_i=-e_i$,
impossible for $e_i=\pm1$).
The leg-swap $\sigma$ fixes the four configurations whose leg bonds
are equal ($e_0=e_2$): $(+,+,+)$, $(+,-,+)$, $(-,+,-)$, $(-,-,-)$.}
\label{tab:burnside_iso}
\begin{ruledtabular}
\begin{tabular}{llr}
Element $g$ & Action on $(e_0,e_1,e_2)$ & $|\mathrm{Fix}(g)|$ \\
\hline
$e$      & $(e_0,\,e_1,\,e_2)$       & 8 \\
$T$      & $(-e_0,-e_1,-e_2)$        & 0 \\
$I$      & $(-e_2,-e_1,-e_0)$        & 0 \\
$\sigma$ & $(e_2,\,e_1,\,e_0)$       & 4 \\
\hline
\multicolumn{2}{r}{Burnside: $(8+0+0+4)/4$} & $=\mathbf{3}$ \\
\end{tabular}
\end{ruledtabular}
\end{table}

\noindent
The isosceles geometry yields \textbf{3 symmetry classes},
one more than the equilateral case (which gave 2 under
$\langle R,T,I\rangle$). Burnside's lemma yields three orbits listed in Table~\ref{tab:orbits_iso}.
They are distinguished by two quantities: i) the magnitude of the winding number $|W|$, and ii) the
\emph{defect position}, position of the minority bond.

\paragraph{Orbit~I — pure circulation ($|W|=3$, size 2).}
The fully aligned configurations $(+,+,+)$ and $(-,-,-)$ form a
$T$-doublet.
The leg-swap $\sigma$ fixes both members, since all bonds carry equal
currents and exchanging the two legs changes nothing.
The stabilizer $\{e,\sigma\}$ reflects this residual symmetry.

\paragraph{Orbit~II — base-bond defect ($|W|=1$, size 2).}
The configurations $(+,-,+)$ and $(-,+,-)$ have their minority bond
on the base.
$\sigma$ fixes $(+,-,+)$: the two legs both carry $e=+1$, so
swapping them leaves the configuration unchanged.
The stabilizer is again $\{e,\sigma\}$.
$T$ and $I$ exchange the two members.

\paragraph{Orbit~III — leg-bond defect ($|W|=1$, size 4).}
The four configurations $\{(+,+,-),(-,+,+),(+,-,-),(-,-,+)\}$
have their minority bond on one of the two legs.
$\sigma$ maps $(+,+,-)\mapsto(-,+,+)$: it moves the defect from
leg~$e_0$ to leg~$e_2$.
Since no element of $G_\mathrm{iso}$ fixes any member,
the stabilizer is trivial $\{e\}$, and the orbit has the
maximal size $|G_\mathrm{iso}|/1 = 4$.

\subsection{Non-reciprocity conditions for the isosceles ring}

For the pure-circulation class (Orbit~I), both $T$ and $I$ map
$(+,+,+)\leftrightarrow(-,-,-)$.
Since $G_\mathrm{iso} = \langle T,I\rangle$ and both generators
protect this degeneracy, the non-reciprocity condition is the same
as for the equilateral case: \emph{both $T$ and $I$ must be broken}.

However, the isosceles geometry introduces an important simplification.
The leg-swap $\sigma=IT$ fixes the pure-circulation states
[$\sigma(+,+,+)=(+,+,+)$], so $\sigma$ plays no role in protecting
the Orbit~I degeneracy.
The non-reciprocity condition depends only on $T$ and $I$, which
remain independent physical mechanisms (Zeeman field for $T$,
Rashba SOC for $I$), exactly as in the equilateral case.

\begin{figure}[h]
    \centering    \includegraphics[width=0.4\textwidth]{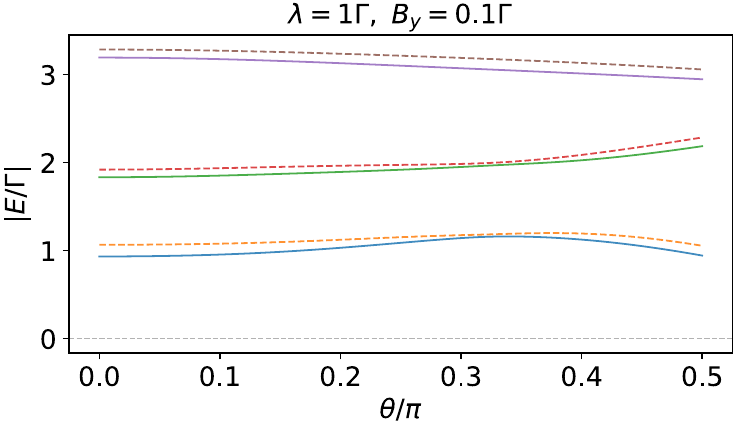}
    \caption{ ABS energy as a function of $\theta$ at $b_y=0.1\Gamma$ and $\lambda=\Gamma$. The energy splittings are enhanced when driven away from $\theta=\pi/3$. }
    \label{fig:isoEabs}
\end{figure}

The new feature of the isosceles ring is the \emph{defect-position}: Orbit~II (base defect) and Orbit~III (leg defect)
are genuinely distinct even with full $G_\mathrm{iso}$ symmetry intact, and their energies can differ without breaking any symmetry. This spectral splitting is a purely geometric effect.
Indeed, in Fig.~\ref{fig:isoEabs}, we show the ABS energies at different $\theta$. The ABS energies splitting are enhanced as we are away from $\theta=\pi/3$. This leads to the further splitting of the nonreciprocal response energy in Fig.~\ref{fig:cir} and results in a circulatory band.


\section{Classification of the $n=2$ ring: the Josephson diode}
\label{app:n2}

The $n=2$ ring is the simplest case and maps directly onto the
Josephson diode. Its classification reveals a key structural difference from $n=3$: the $W=0$ sector is topologically accessible (even $n$), and the relevant configurations for the diode effect live in a $T$-\emph{invariant} orbit.

\subsection{Configuration space}

With $n=2$ vertices and 2 bonds, the configuration space has
$|\mathcal{C}|=2^2=4$ elements.
The winding number $W(\bm{e})=e_0+e_1\in\{-2,0,+2\}$ satisfies
$W\equiv0\pmod{2}$ (even $n$), so $W=0$ is allowed.
The four configurations and their physical meaning are:

\begin{table}[h]
\centering
\caption{All $n=2$ configurations and their physical character.
Bond~0 runs from vertex~0 to vertex~1 (left to right);
bond~1 runs from vertex~1 back to vertex~0.
$W=\pm2$ corresponds to loop currents with no net transport;
$W=0$ corresponds to net transport between the two terminals.}
\label{tab:n2_configs}
\begin{ruledtabular}
\begin{tabular}{ccrll}
$e_0$ & $e_1$ & $W$ & Physical character & Orbit \\
\hline
$+1$ & $+1$ & $+2$ & CW loop current, no net transport & $\mathcal{P}$ \\
$-1$ & $-1$ & $-2$ & CCW loop current, no net transport & $\mathcal{M}$ \\
$+1$ & $-1$ & $0$  & Net current $L\to R$ (diode forward) & $\mathcal{Q}$ \\
$-1$ & $+1$ & $0$  & Net current $R\to L$ (diode reverse) & $\mathcal{Q}$ \\
\end{tabular}
\end{ruledtabular}
\end{table}

\subsection{Symmetry group}

The symmetry group $G=\langle R,T,I\rangle$ has a key algebraic simplification for $n=2$: the rotation $R$ (by $\pi$) satisfies $R^2=e$, and inversion $I$ coincides with $TR$:
\begin{equation}
  I = TR = RT \quad (n=2\text{ only}).
  \label{eq:n2_I_eq_TR}
\end{equation}
This means $I$ is not an independent generator for $n=2$. The group generated
by $\{R,T,I\}$ is simply $\langle R,T\rangle\cong\mathbb{Z}_2\times\mathbb{Z}_2$
of order~4, with elements $\{e,R,T,I=TR\}$.

The transformation laws are:
\begin{align}
  W(R\bm{e}) &= W(\bm{e}),\quad
  W(T\bm{e}) = -W(\bm{e}),\quad
  W(I\bm{e}) = -W(\bm{e}).
\end{align}
Since $I=TR$ for $n=2$, spatial inversion and time-reversal followed by
rotation are the same operation.
This is in contrast to $n=3$, where $I$ and $T$ are independent.

\subsection{Burnside enumeration}

\begin{table}[h]
\centering
\caption{Fixed-point counts for $G=\langle R,T,I\rangle\cong\mathbb{Z}_2\times\mathbb{Z}_2$
($|G|=4$) acting on the $n=2$ configuration space.
$T$ fixes nothing (requires $e_i=-e_i$: impossible).
$I=TR$ fixes the $W=0$ configurations $(+,-)$ and $(-,+)$
because $I(+,-)= [-(-),-( +)]=(+,-) $ and likewise for $(-,+)$.
Burnside: $8/4=2$ orbits.}
\label{tab:burnside_n2}
\begin{ruledtabular}
\begin{tabular}{llr}
Element $g$ & Fixed-point condition & $|\mathrm{Fix}(g)|$ \\
\hline
$e$     & all configurations & 4 \\
$R$     & $e_0=e_1$: $(+,+)$ and $(-,-)$ & 2 \\
$T$     & $e_i=-e_i$: impossible & 0 \\
$I=TR$  & $e_0=-e_1$: $(+,-)$ and $(-,+)$ & 2 \\
\hline
\multicolumn{2}{r}{Burnside: $8/4$} & $=\mathbf{2}$ \\
\end{tabular}
\end{ruledtabular}
\end{table}

Burnside's lemma yields \textbf{two orbits}, listed in
Table~\ref{tab:n2_orbits}.

\begin{table}[h]
\centering
\caption{Orbit classification for the $n=2$ ring under
$G=\langle R,T,I\rangle\cong\mathbb{Z}_2\times\mathbb{Z}_2$.
Orbit $\{\mathcal{P},\mathcal{M}\}$ contains the loop-current configurations;
orbit $\mathcal{Q}$ contains the transport configurations.
All $2^2=4$ configurations accounted for: $2+2=4$.}
\label{tab:n2_orbits}
\begin{ruledtabular}
\begin{tabular}{cccll}
Orbit & $W$ & Size & $\mathrm{Stab}$ & Members \\
\hline
$\{\mathcal{P},\mathcal{M}\}$ & $\pm2$ & 2
  & $\{e,R\}$
  & $(+,+)$,\;$(-,-)$ \\
$\mathcal{Q}$                 & $0$    & 2
  & $\{e,I\}$
  & $(+,-)$,\;$(-,+)$ \\
\end{tabular}
\end{ruledtabular}
\end{table}

\paragraph{Orbit $\{\mathcal{P},\mathcal{M}\}$ — loop currents ($|W|=2$).}
The configurations $(+,+)$ and $(-,-)$ represent clockwise and
counterclockwise loop currents.
$R$ fixes both ($R(+,+)=(+,+)$: swapping two equal bonds changes nothing),
giving stabilizer $\{e,R\}$ and orbit size~$2$.
$T$ maps $\mathcal{P}\leftrightarrow\mathcal{M}$, and likewise $I=TR$.
These configurations carry no net current between terminals and are
irrelevant for the diode effect.

\paragraph{Orbit $\mathcal{Q}$ — diode transport ($W=0$).}
The configurations $(+,-)$ and $(-,+)$ represent net current flowing
$L\to R$ and $R\to L$ respectively.
$I=TR$ fixes both members: $I(+,-)=(-(-), -(+))=(+,-)$, so $\mathrm{Stab}=\{e,I\}$
and orbit size~$2$.
Crucially, $T$ also maps \emph{within} $\mathcal{Q}$:
$T(+,-)=(-,+)\in\mathcal{Q}$.
This means $\mathcal{Q}$ is a \emph{$T$-invariant orbit}.
Similarly, $R$ maps within $\mathcal{Q}$: $R(+,-)=(-,+)\in\mathcal{Q}$.

\subsection{Non-reciprocity condition: the Josephson diode}

The diode effect requires distinguishing $(+,-)$ from $(-,+)$ within orbit
$\mathcal{Q}$ making the forward current $I_c^+$ different from the
reverse current $|I_c^-|$.
Since $T$, $R$, and $I$ all map \emph{within} $\mathcal{Q}$, the two members
are related by every generator of $G$, and \emph{all three symmetries must be
broken} to split them:
\begin{equation}
  \text{Diode effect}
  \;\Longleftrightarrow\;
  T\;\text{broken \emph{and}}\;R\;\text{broken.}
  \label{eq:diode_condition}
\end{equation} 
Physically, breaking $T$ requires a magnetic flux or Zeeman field, and
breaking  the spatial symmetry requires a
geometrically or electrostatically asymmetric junction.
This is precisely the Josephson diode recipe observed experimentally using magnetic field plus asymmetric junction.



\end{document}